\documentclass[twocolumn,showpacs,superscriptaddress,preprintnumbers,amsmath,amssymb,epsfig,floatfix,prl]{revtex4-1}
\usepackage{epsfig}
\usepackage{graphicx}
\usepackage{dcolumn}
\usepackage{color}
\usepackage{natbib}  
\usepackage{hyperref}
\usepackage{breakurl}
\hypersetup{colorlinks=true, citecolor=blue, urlcolor=blue, linkcolor=blue}
\usepackage{bm}
\usepackage{tabularx}
\newcolumntype{L}[1]{>{\raggedright\arraybackslash}p{#1}}
\newcolumntype{C}[1]{>{\centering\arraybackslash}p{#1}}
\newcolumntype{R}[1]{>{\raggedleft\arraybackslash}p{#1}}
\begin{document}
\title{Photocatalytic Activity of Phosphorene Derivatives: Coverage, Electronic, Optical and Excitonic properties}
\author{Srilatha Arra}
\affiliation{Department of Chemistry, Indian Institute of Science Education and Research, Pune 411008, India}
\author{K. R. Ramya}
\altaffiliation{Present address: Max Planck Institute for Chemical Physics of Solids, N\"othnitzer Strasse 40, 01187 Dresden, Germany}
\affiliation{Department of Physics, Indian Institute of Science Education and Research, Pune 411008, India}
\author{Rohit Babar}
\affiliation{Department of Physics, Indian Institute of Science Education and Research, Pune 411008, India}
\author{Mukul Kabir\thanks{Corresponding author}}
\email{Corresponding author: mukul.kabir@iiserpune.ac.in} 
\affiliation{Department of Physics, Indian Institute of Science Education and Research, Pune 411008, India}
\affiliation{Centre for Energy Science, Indian Institute of Science Education and Research, Pune 411008, India}
\date{\today}

\begin{abstract} 
In the  context of two-dimensional metal-free photocatalyst, we investigate the electronic, optical and excitonic properties of phosphorene derivatives within first-principles approach. While two-dimensional phosphorene does not catalyze the complete water splitting reactions, O, S, and N coverages improve the situation drastically, and become susceptible to catalyze the complete reaction at certain coverages. We find that for all these dopants, 0.25 -- 0.5 ML coverages are thermodynamically more stable, and does not introduce midgap defect states and the composite systems remain semiconducting along with properly aligned valance and conduction bands. Further, within visible light excitation, the optical absorption remain very high 10$^5$ cm$^{-1}$ in these composite systems, and the fundamental optical anisotropy of phosphorene remains intact. We  also investigate the effect of layer thickness through bilayer phosphorene with oxygen coverages. Finally we investigate the excitonic properties in these composite materials that are conducive to both redox reactions.  The present results will open up new avenues to take advantage of these metal-free phosphorene derivatives toward its outstanding potential in photocatalysis.  
\end{abstract}
\maketitle

\section{Introduction}
Hydrogen is a promising energy carrier, which can be used along with oxygen in a fuel cell to convert chemical energy into electricity.~\citep{10.1063/1.1878333} Although highly abundant, hydrogen occurs in chemical compounds such as hydrocarbons and water, which must be chemically transformed to produce H$_2$. Further, water splitting H$_2$O = 1/2O$_2$ + H$_2$ is an uphill reaction, and the change is Gibbs free energy is 237 kJ mol$^{-1}$, which according to Nernst equation corresponds to 1.23 eV per electron transfer.~\citep{10.1038/238037a0,10.1021/cr1001645,10.1021/jz502646d}

In the presence of light, semiconductors may catalyze the H$_2$O oxidation reaction if the incident photons have sufficient energy to generate electron-hole ($e-h$) pairs. While the hole participates in the oxidation reaction,  H$_2$O + 2$h$ = 1/2O$_2$ + 2H$^+$, the excited electron drives the consequent hydrogen reduction reaction 2H$^+$ + 2$e$ = H$_2$. Since the discovery of photocatalytic water splitting on TiO$_2$, numerous semiconducting materials have been proposed.~\citep{10.1038/238037a0,10.1021/cr1001645,10.1021/jz502646d}  Materials that are in focus for the last 40 years consist of $d^0$ transition metals or the post transition metals with  $d^{10}$ electronic configuration along with the group VA and VIA ions.~\citep{10.1038/238037a0,10.1021/cr1001645,10.1021/ja0269643, 10.1021/ja048296m, ANIE:ANIE200500314, 10.1038/440295a, 10.1021/jp0656532, ANIE:ANIE201204675} However, the practical use of these materials are limited by their inability to absorb visible light, and fast recombination leading to insufficient quantum efficiency. Nanoscale materials with tuneable electronic structure have attracted enormous attention to improve performance.~\citep{Lin2011209} For example, solar-to-hydrogen conversion efficiency is increased up to 16\% for TiO$_2$ nanotubes, while that of the bulk rutile phase is only 2\%.~\citep{10.1021/jp064020k} Further, the two-dimensional materials pose several advantages due to their tuneable electronic and optical properties. Moreover, they maximize the active surface area, and reduce electron-hole recombination through minimizing the distance that the photogenerated carriers travel before reaching catalyst-water interface, where the reaction takes place. Recent experiments corroborate these facts, and single to few layer SnS$_2$, ZnSe, WS$_2$, SnSe, and SnS show order-of-magnitude improvement in the efficiency compared to their bulk counterparts.~\citep{ANIE:ANIE201204675,10.1038/ncomms2066,10.1038/nmat3700,AENM:AENM201300611}

Metal-free photocatalysts are attracting more attention in recent times due to their corrosion free and non-toxic nature along with better processability.~\citep{10.1038/nmat2317, Liu970, 10.1021/acs.chemrev.6b00075} The quantum efficiency for two-dimensional graphitic-C$_3$N$_4$ is found to be as high as 26.5\% under visible light illumination with $\lambda >$ 395 nm.~\citep{ANIE:ANIE201403375}  Another 2D material, phosphorene, has attracted enormous scrutiny due to its extraordinary electronic and optical properties; and envisioned in electronic, optoeletronic and spintronic applications.~\citep{10.1038/nnano.2014.35, 10.1063/1.4868132, 10.1021/nn501226z, 10.1038/natrevmats.2016.61, 10.1021/acs.jpcc.6b05069} Further, anisotropic band structure with high carrier mobility, tuneable quasiparticle gap between 0.3 to 2.2 eV,~\citep{10.1038/nnano.2014.35, 10.1021/nn501226z, 10.1038/ncomms5475, 10.1038/nnano.2015.71} and excellent photon absorption ability in the ultraviolet, visible and near infrared region~\citep{10.1038/nnano.2015.112} indicate phosphorene to be a good photocatalyst and deserves concerted attention.~\citep{C5EE03732H} While pristine single layer phosphorene (SLP) has been found to be incapable of catalyzing the complete water decomposition reaction, under the influence of increased pH 8.0, the  SLP is predicted to catalyze both the half-reactions.~\citep{10.1021/jp508618t}  Following these, phosphorene based heterostructures and composites have been investigated, and found to be a promising route to design new materials, which could prevent electron-hole recombination in addition to tuning the band edges.~\citep{10.1021/jp508618t, 10.1021/jp505257g, 10.1021/acs.jpclett.5b00976}  While MoS$_2$/phosphorene and WS$_2$/phosphorene 2D heterostructures are not encouraging,~\citep{10.1021/jp505257g, 10.1021/acs.jpclett.5b00976} TiO$_2$@phosphorene show increased photo-absorption in 250-1200 nm, and concurrent photocatalytic activity.~\citep{10.1038/srep08691} Further, Au-nanoparticle decorated on phosphorene show encouraging photocatalytic activity.~\citep{10.1021/acs.nanolett.5b02895}

However, due to the presence of lone-pair electrons, the single and few-layer phosphorene is not air-stable, and degrades via oxidation.~\citep{10.1021/nl5032293, 10.1038/ncomms10450} A protective Al$_2$O$_3$ or PMMA coating is found to preserve the intrinsic properties, and are useful for electronic and optoelectronic devices.~\citep{10.1038/nnano.2015.71,10.1021/nl5032293, 10.1038/ncomms10450} In contrast, here we utilize the reactive nature of phosphorene, and investigate the electronic and optical properties of surface absorbed O, N, and S in the context of photocatalysis. Indeed, it is reported recently that hydroxyl-functionalized black phosphorus nano-sheets are excellent photocatalyst for visible light ($\lambda >$ 420 nm) H$_2$ evolution with better activity than graphitic-C$_3$N$_4$.~\citep{ADMA:ADMA201605776} Here, we find that 0.25 -- 0.5 ML coverages are thermodynamically most stable, and the band edges align to water redox potentials. Further, the absorption coefficient is found to be very high, 10$^5$ cm$^{-1}$ in these derivatives. Finally, the excitonic properties are studied and discussed in the context of photocatalysis.

\begin{figure*}[t!]
\begin{center}
\includegraphics[angle=90,scale=0.55]{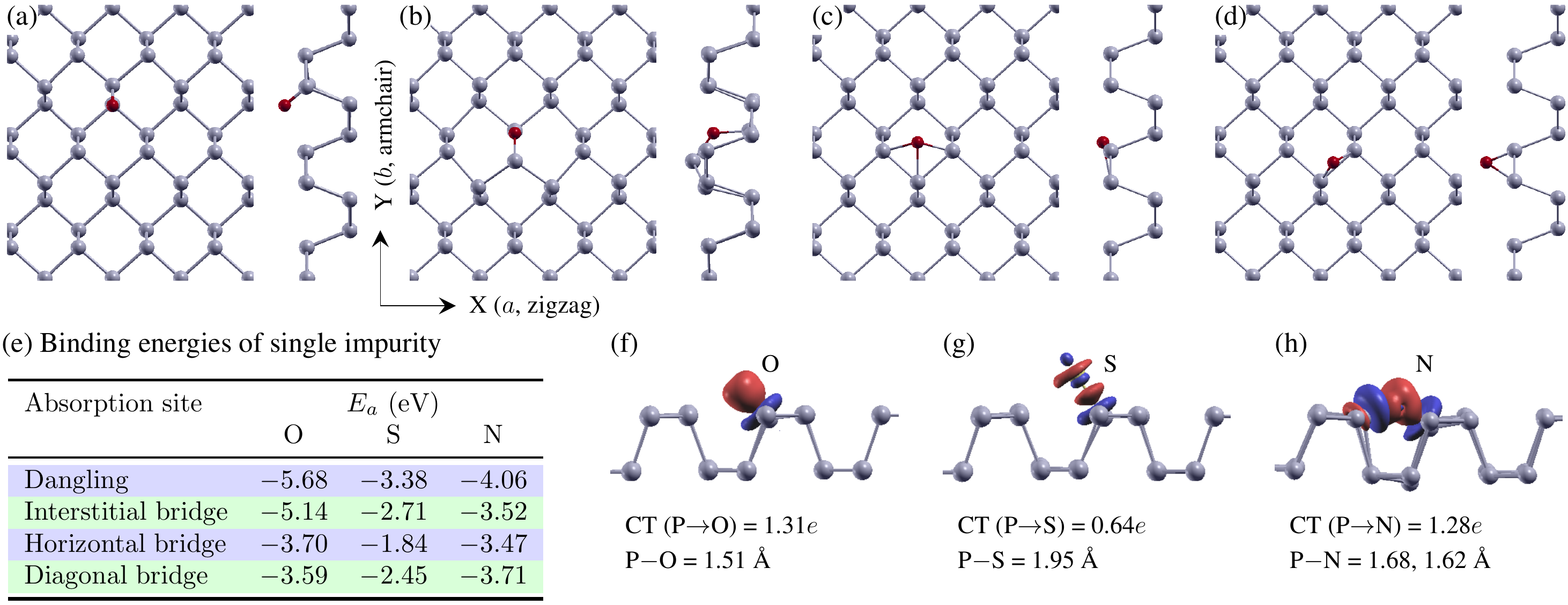}
\vskip -2.8cm
\caption{Different absorption sites on single layer phosphorene, (a) Dangling (b) interstitial bridge (c) horizontal bridge, and (d) diagonal bridge. (e) Binding energies of single impurity are tabulated, which indicate the dangling site absorption to be preferable for all impurities.  Bonding charge density is calculated as $\Delta \rho(r) = \rho_{\rm SLP+X}(r) - \rho_{\rm SLP}(r) - \rho_{\rm X}(r)$, and are shown in (f), (g), and (h) for O, S, and N, respectively. Blue (red) color indicate electron depletion (accumulation). The charge densities are shown for 0.05 $e$/\AA$^3$ isodensity. Bader population analysis of the charge density indicate P$\rightarrow$X charge transfer (CT), and the impurity bonding is linearly correlated with the charge transfer. 
}
\label{figure1}
\end{center}
\end{figure*}

\section{Computational Details}
Calculations were based on the first-principles density functional theory within the projector augmented wave formalism, and the wave functions were expanded in the plane wave basis with 500 eV cutoff.~\citep{PhysRevB.47.558, PhysRevB.54.11169, PhysRevB.50.17953} The exchange correlation energy was described with Perdew-Burke-Ernzerhof (PBE) functional form of generalized gradient approximation.~\citep{PhysRevLett.77.3865} All the structures were fully optimized including volume until the force components were less than 0.01 eV/\AA~ threshold. At least 15 \AA~ of vacuum space was adopted along the z-axis to avoid the periodic interaction.  Absorption energies of P$_x$X$_y$ (X=O, N, S) were calculated by $E_a = \frac{1}{y} [E({\rm P}_x{\rm X}_y) - E({\rm P}_x) - y\mu_{\rm X}]$,  where $x$ and $y$ are the number of P and X atoms in cell used in the calculation. $E({\rm P}_x{\rm X}_y)$ and   $E({\rm P}_x)$ are the energies of SLP with and without X-absorption, while $\mu_{\rm X}$ is the chemical potential of X. Single atom absorption was studied with a 4$\times$3 supercell, and a 8$\times$8$\times$1 $\Gamma$-centred Monkhorst-Pack $k$-grid was used to sample the Brillouin-zone.~\citep{PhysRevB.13.5188}  Energetics of the structures with varied X-coverage $\Theta$ (1, 0.5, 0.25, 0.125 and 0.0625 monolayer, ML) were studied with a 4$\times$4 supercell, along with 8$\times$8$\times$1 $k$-grid.  Further, the electronic structure is calculated within the computationally expensive  Heyd-Scuseria-Ernzerhof (HSE06) hybrid exchange-correlation functional,~\citep{10.1063/1.1564060, 10.1063/1.2404663} and depending on the coverage a minimal (super)cell were used.  For 1, 0.5, 0.33, 0.25, 0.125 and 0.0625 ML coverages, we used 1$\times$1, 2$\times$2,  2$\times$3, 4$\times$1, 4$\times$2,  and 4$\times$4 supercell with commensurate $k$-meshes,  17$\times$17$\times$1, 8$\times$8$\times$1, 8$\times$6$\times$1, 4$\times$17$\times$1,  4$\times$8$\times$1, and 4$\times$4$\times$1, respectively. We incorporated the van der Waals interaction within DFT-D3 method along with Becke-Jonson damping during structural optimization.~\citep{1.3382344, JCC21759}

In the context of photocatalysis the energy positions of the valence band maximum (VBM) and the conduction band minimum (CBM) are the key. To efficiently use the photogenerated electrons and holes, the CBM should be higher (more positive) than the H$^+$/H$_2$ reduction potential, and the VBM should be lower (more negative) than the oxidation potential of O$_2$/H$_2$O. These are $-$4.44 and $-$5.67 eV, respectively, with respect to the absolute vacuum level. Alternately, the water redox potentials are referenced with respect to the normal hydrogen electrode (NHE),   H$^+$/H$_2$ to 0.0 eV and O$_2$/H$_2$O to 1.23 eV.  The standard approach was employed to reposition the band edges with respect to the vacuum level $V(\infty)$, which is the electrostatic potential in the vacuum region far from the 2D material.

\section{Results and Discussion}
We begin our discussion with pristine SLP in the photocatalytic context. SLP has a puckered honeycomb structure resulting from covalently bonded $sp^3$ phosphorus atoms along with a pair of lone-pair electrons. The quasiparticle gap is experimentally measured to be in 2.05$-$2.20 eV range through scanning tunnelling spectroscopy and photo-luminescence excitation spectroscopy.~\citep{10.1038/nnano.2015.71,10.1021/nl502892t} While it is known that the conventional exchange-correlational functionals underestimate the gap (PBE: 0.95 eV),~\citep{10.1021/acs.jpcc.6b05069} the HSE06 hybrid functional (1.4 -- 1.7 eV)\citep{10.1038/ncomms5475,10.1021/nl5021393,10.1021/jp506881v} and many-body perturbation based GW (2.00 -- 2.25)\citep{PhysRevB.89.235319,PhysRevLett.115.066403} calculations improve the results. As the GW calculations are computationally very expensive for the systems that we have investigated in this paper, we resort to HSE06 functional throughout the present investigation for electronic structure. The PBE-D3 optimized lattice parameters, $a$ (zigzag) = 3.29 \AA\ and $b$ (armchair) = 4.45 \AA\ compare well with the bulk black phosphorous.~\citep{Brown} The calculated HSE06 gap of 1.41 eV is in agreement with the previous calculations.~\citep{10.1038/ncomms5475} 
While the CBM of pristine SLP is more negative than H$^+$/H$_2$ redox potential, the VBM is not more positive than O$_2$/H$_2$O oxidation potential (1.23 eV with respect to NHE). Thus, at ambient conditions, SPL is capable of catalyzing only half of the reaction, while in alkaline medium it catalyzes both the half-reactions. These results are in agreement with a previous report, which also indicate that strain engineering may improve the photocatalytic properties.~\citep{10.1021/jp508618t}

\subsection{Isolated impurities}
Next we investigate the absorption O, S, and N  with SLP, which naturally reacts with these impurities due to the presence of lone-pair electrons. Possible absorption sites are shown in Figure~\ref{figure1}(a)-(d). These impurities are chemisorbed on SLP with strong binding energies [Figure~\ref{figure1}(e)], and  are absorbed in a dangling configuration [Figure~\ref{figure1}(a)]. Absorbed oxygen bonds with a single phosphorus atom with a short P$-$O bond of 1.51 \AA\, and very strong binding energy of -5.68 eV. According to Pauling electronegativity of P and O atoms, the P$-$O is polar covalent, which is evident from the bonding charge density and Bader population analysis [Figure~\ref{figure1}(f)], which indicate a large P $\rightarrow$ O charge transfer of 1.31$e$. These results are in agreement with an earlier report.~\cite{PhysRevLett.114.046801} The O absorption at the interstitial bridge site with P$-$O$-$P bond is a metastable configuration with 0.54 eV higher energy. Here O lies between the two half-layers forming two P$-$O bonds, 1.65 and 1.69 \AA\ and 130$^\circ$ P$-$O$-$P bond angle. The overall results are in agreement with  recent X-ray photoemission spectroscopic measurement, which indicates the abundance of P$-$O bonds over  P$-$O$-$P bonds.~\citep{ADMA:ADMA201605776}
Similarly, sulfur is absorbed at the dangling site, however, the binding energy is much smaller than O absorption. This can be explained as the electronegativity of P and S atoms are comparable, and thus the corresponding P$-$S bond is nonpolar and longer (1.95 \AA). This is further corroborated by the bonding charge density, which indicates nonpolar covalent P$-$S bonding, and Bader analysis indicate only 0.64$e$ charge transfer from P to S [Figure~\ref{figure1}(g)]. 
Similar to O and S, nitrogen is absorbed at the dangling  site, however which forms two P--N bonds with 1.62 and 1.68 \AA\ distances.  As the electronegativity of N is intermediate of O and S, the chemisorption energy is also found to be intermediate [Figure~\ref{figure1}(e)], and the resulting bond is polar covalent [Figure~\ref{figure1}(h)]. The corresponding P $\rightarrow$ N charge transfer is found to be 1.28$e$, which is commensurate with bonding. Note that for all impurities, absorption at the various  bridge sites [Figure~\ref{figure1}(b)-(d)] are found to be much higher in energy [Figure~\ref{figure1}(e)]. 

\begin{figure*}[t!]
\begin{center}
\includegraphics[angle=90,scale=0.38]{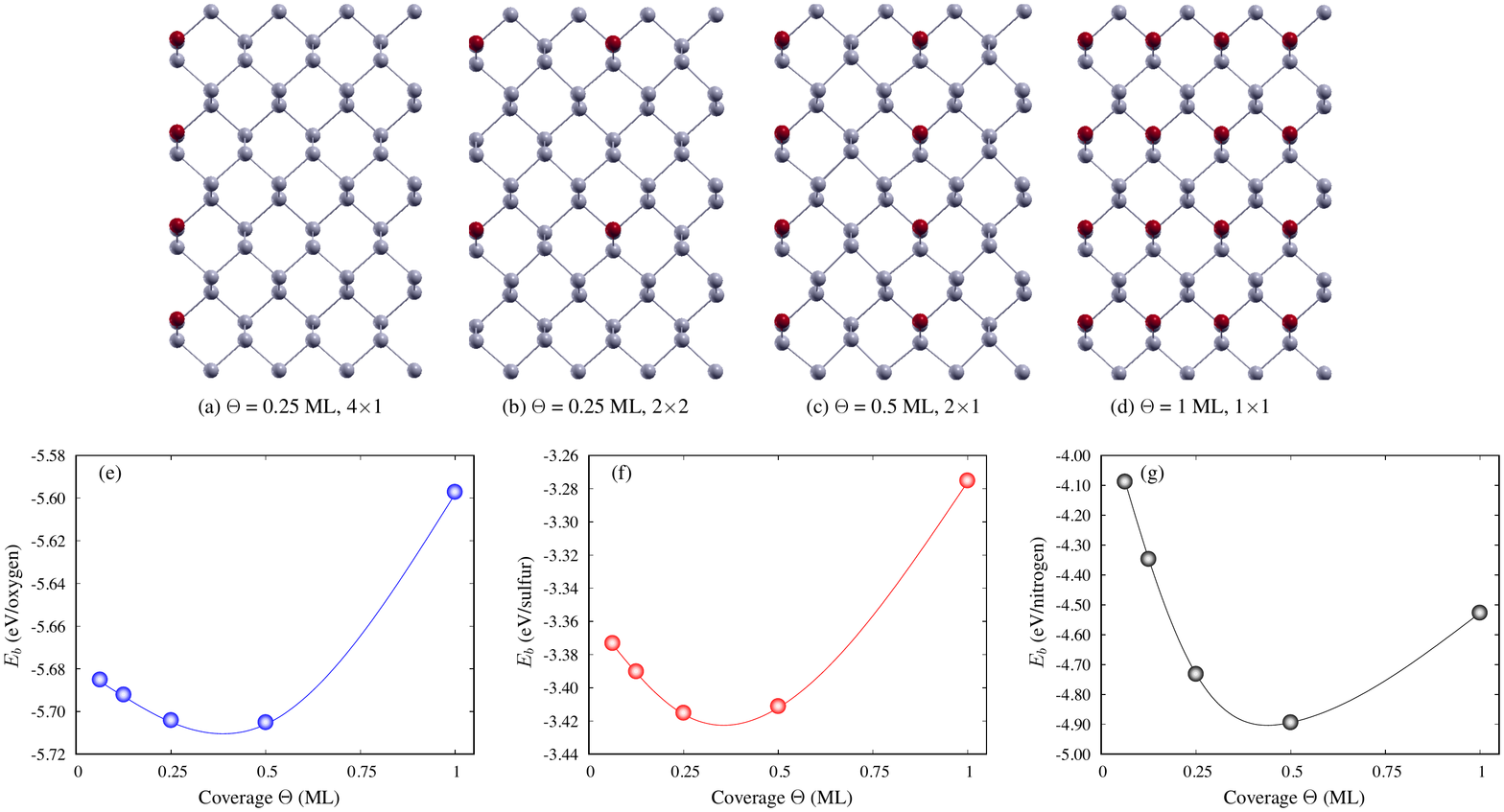}
\vskip -1.0cm
\caption{Representative configurations for impurity coverages shown for oxygen and sulfur, (a) $\Theta$ = 0.25 ML -- 4$\times$1, (b) $\Theta$ = 0.25 ML -- 2$\times$2, (c) $\Theta$ = 0.5 ML -- 2$\times$1, and (d) $\Theta$ = 1 ML -- 1$\times$1. Other configurations are shown in Supporting Information.  Impurities are absorbed at the dangling site.  (e)-(g) Calculated binding energy with varied impurity coverage $\Theta \leqslant$ 1. Regardless the impurity type, the optimal coverage is found in the 0.25 -- 0.5 ML range, which optimize the impurity--phosphorene bonding,  Coulomb interaction between the impurities and the induced strain in SLP lattice.
}
\label{figure2}
\end{center}
\end{figure*}

\subsection{Impurity coverage}
Having discussed the individual impurity absorption, we now turn our attention to impurity coverage $\Theta$ $\leqslant$ 1 ML, and investigate their electronic structure, band alignment  and optical properties. First, we calculate the binding energy with varied impurity coverage, which is a tricky task as there might be several possible configurations for a particular coverage. Here, we consider only one sided absorption, as if the SLP is placed on a non-interacting substrate.  Impurities are chemisorbed at the most stable dangling site [Figure~\ref{figure1}(e)].  Representative configurations for different coverages are shown in Figure~\ref{figure2}(a)-(d) for $\Theta$ = 0.25, 0.5 and 1 ML. Other configurations are shown in Supplementary Information. Apart from the chemical interaction between the impurity and SLP, the impurity binding energy will be affected by varied coverage through Coulomb interaction between the impurities, and the strain developed in the SLP lattice due to impurity loading. 
For oxygen, the optimal coverage is found to be between 0.25 to 0.5 ML, and beyond which calculated binding energy decreases [Figure~\ref{figure2}(e)]. While the 4$\times$1 configuration is found to be most stable for 0.25 ML oxygen, the  2$\times$2 impurity configuration lies only 5 meV/O higher in energy. In contrast, the 1$\times$4 configuration lies much higher (113 meV/O) for the same coverage. For 0.5 ML oxygen, 2$\times$1 configuration is found to be the ground state, which is 37 and 93 meV/O lower in energy than the corresponding $c$(2$\times$2), and 1$\times$2  impurity configurations, respectively. Further, when the coverage is increased to 1 ML, oxygen binding is destabilized by 82 meV/O compared to isolated impurity absorption (Figure~\ref{figure1}). It should be noticed here that the P$-$O bond length does not change ($\sim$ 0.1\%) due to increased oxygen loading (Supporting Information). In contrast, for the most stable configurations with increasing oxygen coverage, the lattice parameters increases by 0.8\% and 2.7\% along $a$ (zigzag) and $b$ (armchair) directions for 1 ML compared to the pristine SLP. 

The picture is qualitatively similar for sulfur, and the binding is maximum for  0.25 -- 0.5 ML coverage [Figure~\ref{figure2}(f)].  For 0.25 ML sulfur, the 2$\times$2 configuration is preferred, while the competing 4$\times$1 is 26 meV/S higher. With increasing coverage at 0.5 ML, the 2$\times$1 configuration is the ground state, which is 35 meV/S lower in energy than the corresponding $c$(2$\times$2) configuration. Beyond 0.5 ML coverage, sulfur binding decreases rapidly.  
The change in P$-$S bond length and $a$ do not show any noticeable change till 0.5 ML coverage, whereas $b$ increases monotonically by 4.3\% (Supporting Information). However, at 1 ML coverage, the P$-$S bonds are increased by 0.05 \AA, while $a$ and $b$ lattice parameters are increased by  2.9 and 6.7\%, respectively, compared to pristine SLP. Thus, induced strain is larger for sulfur compared to oxygen coverage.

The binding energy is found to be maximum around 0.5 ML N-coverage [Figure~\ref{figure2}(g)]. For 0.5 ML coverage, the 2$\times$1 configuration is found to be the most stable over other configurations with binding energy 23 meV and 340 meV/N lower than the $c(2\times2)$ and 1$\times$2 configurations, respectively.  Beyond 0.5 ML coverage,  $E_b$ decreases much rapidly with varied coverage unlike oxygen and sulfur. Compared to 0.5 ML coverage, the calculated $E_b$ increases by 367 and  547 meV/N for 1 ML and 0.125 ML coverages, respectively. In contrast, such increase is found to be much smaller (below 150 meV/dopant) for O and S coverages [Figure~\ref{figure2}(e)-(g)]. We argue this feature to be connected to the lattice strain developed due to N-loading, which we notice to be much different from the cases with O and S doping, where only a tensile strain is developed in both zigzag and armchair direction. In contrast, a tensile and a concurrent compressive strains are developed along zigzag and armchair directions, respectively (Supporting Information), which monotonically increases with increasing coverage.  For 1ML N-coverage, the SLP lattice parameters increase by 3.7\% (tensile strain) along the zigzag direction, while the same decreases by 5.3\% (compressive strain) along the armchair direction. 

It would be interesting to investigate the differential effects of induced strain on the impurity binding. It is known that due to puckered nature along the armchair direction, the SLP can easily accommodate large strain along this direction, while the zigzag direction is relatively stiffer.~\cite{10.1063/1.4885215}  In this regard, we have calculated the strain energy relative to the pristine SLP, $\mathcal{E}_s$ = $[E'({\rm SLP}) - E({\rm SLP})]/{\rm area}$, where $E$ and  $E'$ are the energies of pristine and distorted SLP due to impurity loading. Calculate strain energy increases monotonically with impurity loading, and is much smaller compared the corresponding impurity binding energy (Supporting Information). For any coverage, we find $\mathcal{E}^{\rm O}_s < \mathcal{E}^{\rm S}_s < \mathcal{E}^{\rm N}_s$, which is commensurate with the respective changes in their lattice parameters due to impurity absorption.  For example, while $\mathcal{E}_s$ is comparable for $\Theta$= 0.5 ML O and S coverages (75 and 105 mJ/m$^{-2}$, respectively), the same for N-coverage is much larger to 570 mJ/m$^{-2}$. Furthermore, it is found that impurities repel each other due to Coulomb interaction while the separation fall below a 4 \AA\ distance, and consequently binding energy decreases.  

\subsection{Electronic structure and band alignment}
\begin{figure}[t!]
\begin{center}{
\vskip -2.8cm
\includegraphics[angle=0,scale=0.35]{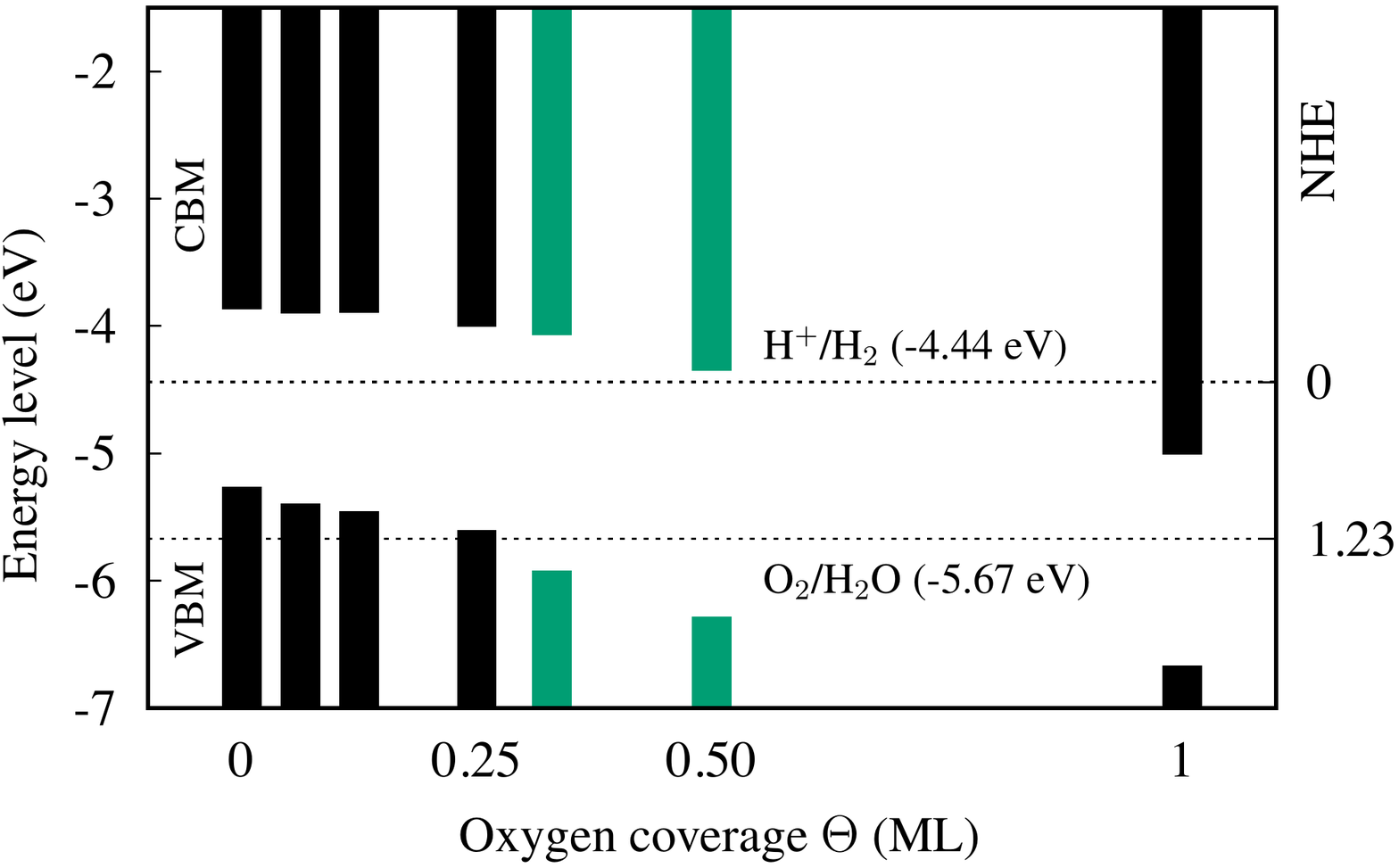}
\vskip -6cm
\includegraphics[angle=0,scale=0.35]{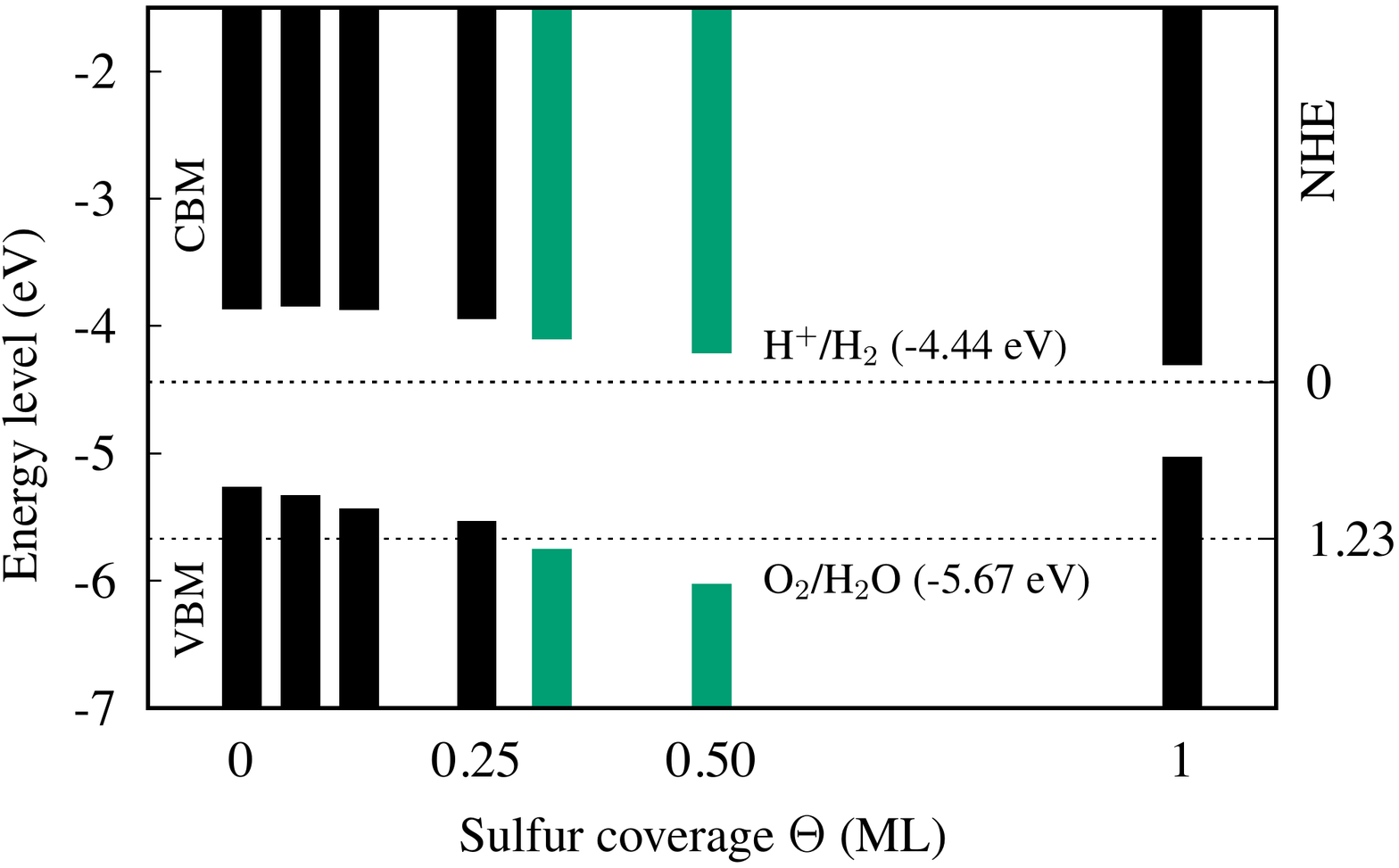}}
\vskip -2.8cm
\caption{Band-edge positions of single layer phosphorene with varied oxygen and sulfur coverage at ambient conditions. These are calculated with hybrid HSE06 exchange-correlation functional and rescaled to the vacuum level and NHE. The redox potentials of H$_2$O is marked with the dashed lines. The band-edge positions for SLP with 0.33 and 0.5 ML oxygen and sulfur coverages are such that they catalyze both the half reactions.  
}
\label{figure3}
\end{center}
\end{figure}

The band structure of pristine SLP has been studied extensively in literature using various theoretical hierarchy and accuracy,~\citep{10.1021/acs.jpcc.6b05069, 10.1038/ncomms5475,10.1021/nl5021393,10.1021/jp506881v, PhysRevB.89.235319,PhysRevLett.115.066403}  and the optical and quasiparticle gaps have been experimentally estimated.~\citep{10.1038/nnano.2015.71,10.1021/nl502892t} As conventional exchange-correlation functionals underestimate the band gap, we use hybrid HSE06 functional throughout the paper. Further, magnitude of the gap, and the corresponding energy positions of VBM and CBM are necessary for photocatalytic applications. In this context, as we have discussed earlier, pristine SLP catalyze only the H$^+$/H$_2$ half of the water oxidation, while O$_2$/H$_2$O reaction is not possible due to improper energy position of VBM (Figure~\ref{figure3}). The calculated gap of 1.41 eV is in good agreement with the previous calculations within similar theoretical hierarchy. \citep{10.1038/ncomms5475,10.1021/nl5021393,10.1021/jp506881v}  Here, we investigate to tune the gap and the corresponding VBM/CBM alignment with varied adatom coverages $\Theta \leqslant 1$ ML. 

We start discussion with varied oxygen coverage, which does not introduce any electronic  states in the gap and the composite system remain semiconducting. Calculated gap shows an overall non-monotonous variation. With increasing oxygen coverage, the gap first increases monotonously to 1.95 eV for 0.5 ML coverage, however which decreases to 1.68 eV for 1 ML oxygen. It is evident from the density of states (DOS) that both VBM and CBM originate from the P-$p_z$ orbital for pristine SLP,  and this picture remains unperturbed with increasing oxygen coverage till 0.5 ML. We do not observe any contribution from oxygen to the band edges. However, for 1 ML coverage, hybridizied P-$p_y$ and O-$p_y$ contribution is observed in addition to P-$p_z$ orbital at the valence band edge. Furthermore, the conduction band edge completely changes its character to P-$p_x$. 

With increasing coverage, the band edges come down in energy (Figure~\ref{figure3}) and the band-edge positions are such that for 0.33 and 0.5 ML become susceptible to catalyze both the water oxidation reactions. We have further investigated the oxidizing power $\Delta_{\rm VBM} = (E_{\rm VBM} + 5.67)$ eV, and reducing power $\Delta_{\rm CBM} = (E_{\rm CBM} + 4.44)$ eV. 
This is the energy measure of how negative (positive) the VBM (CBM) lies with respect to the respective redox potential. A large difference between $\Delta_{\rm VBM}$ and $\Delta_{\rm CBM}$ imbalances the oxidation and reduction reactions, which inhibit further redox reaction.  Thus, similar $\Delta_{\rm VBM}$ and $\Delta_{\rm CBM}$ is required for overall catalytic efficiency. 
In this context, the SLP with 0.33 ML coverage ($\Delta_{\rm VBM}$ = $-$0.26 eV and $\Delta_{\rm CBM}$ = 0.38 eV) is expected to be a better catalyst than that with 0.5 ML coverage ($\Delta_{\rm VBM}$ = $-$0.62 eV and $\Delta_{\rm CBM}$ = 0.10 eV). The case with 0.25 ML coverage is interesting. While the VBM is unfavorable with being slightly positive (0.06 eV) compared to the oxidation potential, the CBM shows preferable alignment.  However, increased pH of the solution will align the VBM in this case, as the redox potentials shift higher in energy by pH$\times$0.059 eV.~\citep{Chakrapani1424} Thus, in a moderate pH 5 solution, both VBM and CBM align preferably with equivalent $\Delta_{\rm VBM}$ = $-$0.24 eV and $\Delta_{\rm CBM}$ = 0.15 eV, which indicate a balanced redox reactions.  Although coverages below 0.25 ML are not thermodynamically preferable [Figure~\ref{figure2}(e)], at moderate pH 5, structures with 0.0625 ML ($\Delta_{\rm VBM}$ = $-$0.03 eV and $\Delta_{\rm CBM}$ = 0.25 eV) and 0.125 ML ($\Delta_{\rm VBM}$ = $-$0.08 eV and $\Delta_{\rm CBM}$ = 0.26 eV) O-coverages are favourable to redox reactions.   

Next we discuss oxygen coverage on bilayer phosphorene (BLP), for which we consider the natural AB stacking that is energetically favorable.  Calculated lattice parameters, within the PBE-D3 functional,  are 3.30 and 4.33 \AA~ along the zigzag and armchair directions, respectively. Calculated interlayer distance of 5.11 \AA~ is in good agreement with  previous quantum Monte Carlo results.~\citep{doi:10.1021} Note that here we have considered one-sided oxygen coverage only. While the lattice parameter along the zigzag direction is not perturbed by O coverage, the same along the armchair direction monotonically increases to 4.48 \AA~ for 1ML coverage. The calculated binding energy with varied coverage show the same qualitative trend as discussed for the SLP [Figure~\ref{figure2}(e)], while the binding is about 50 meV higher in case of BLP. The dangling $d_{\rm P-O}$ does not change upon increasing O loading, which remain at 1.51 \AA.

Compared to pristine SLP, the band gap decreases in pristine BLP to 1.10 eV, calculated within hybrid HSE06 functional, which is in good agreement with the previous reports.~\citep{10.1038/ncomms5475,10.1021/jz500409m} The trend in band gap with increasing O coverage for BLP shows similar qualitative trend as it is discussed for SLP earlier (Supporting Information).  The gap increases with increasing coverage till 0.5 ML to 1.44 eV, however, it decreases with further increase in coverage (1.32 eV for 1 ML).  Similar to the trend observed in SLP, the band edges come down in energy with increasing coverage, and the BLP with 0.5 ML oxygen coverage becomes capable to catalyze both redox reactions with balanced oxidizing and reducing power ($\Delta_{\rm VBM}$ = $-$0.09 eV and $\Delta_{\rm CBM}$ = 0.08 eV). Note that for $\Theta \leqslant$ 0.25 ML coverage, BLP catalyzes only one half of the complete reaction, H$^+$/H$_2$. However, for 0.25 ML O-coverage, BLP  is predicted to catalyze both the half reactions at a moderate pH 4 solution ($\Delta_{\rm VBM}$ = $-$0.08 eV and $\Delta_{\rm CBM}$ = 0.09 eV). 

Motivated by our results with oxygen coverage, we have further investigated the role of sulfur. The calculated binding energy is maximum for 0.25--0.5 ML sulfur coverage, as we have discussed earlier [Figure~\ref{figure2}(f)]. Similar to the trend observed for oxygen coverage, absorbed sulfur does not introduce any midgap states at any coverage $\leqslant$ 1 ML that are studied here, and the composite systems remain semiconducting (Figure~\ref{figure3}). The calculated gap monotonically increases with increasing coverage, 1.5 eV for 0.0625 ML to 1.83 eV for 0.5 ML coverage. However, the calculated gap for 1 ML coverage is severely reduced to 0.74 eV. Interestingly, we find for 0.33 and 0.5 ML sulfur coverages, the band edges align such that these systems become catalytically active. The corresponding oxidation and reduction powers are found to be $\Delta_{\rm VBM}$ = $-$0.09 eV and $\Delta_{\rm CBM}$ = 0.35 eV for 0.33 ML coverage, and the same for 0.5 ML are $-$0.36 and 0.24 eV, respectively.  In contrast, the structures with 0.125 ML ($\Delta_{\rm VBM}$ = $-$0.07 eV and $\Delta_{\rm CBM}$ = 0.28 eV) and 0.25 ML ($\Delta_{\rm VBM}$ = $-$0.17 eV and $\Delta_{\rm CBM}$ = 0.21 eV) S-coverages become favourable to redox reactions at moderate pH 5 only.

The downward shift in band edges after O/S coverage is crucial for phosphorene to exhibit photocatalytic activity (Figure~\ref{figure3}). The shift originate from the change in electron environment at the vacuum-phosphorene interface.~\citep{PhysRevB.30.4874} Due to P$\rightarrow$O/S charge transfer, O and S have surplus electron, and the electron-rich environment at the surface induces a dipole that leads to altered band alignment with respect to the vacuum. Further, the work function $\varphi$, which is the energy difference between the vacuum level $V(\infty)$ and the Fermi energy, show monotonic increase with increasing O/S coverage. This is again due to the dipole formation between SLP and O/S impurity. Calculated $\varphi$ for SLP, 0.0625, 0.125, 0.25, 0.33, 0.5, 1 ML O-coverages are 4.86, 4.93, 5.07, 5.10, 5.50, 5.84 and 6.28 eV, respectively, whereas  $\varphi$ monotonically increases to 5.65 eV for 0.5 ML S-coverage. Such tuneable work function by impurity coverage will have desirable impact on functional device applications.

Next we investigate nitrogen coverage on SLP, and find that the composite system with $\sim$ 0.5 ML coverage is thermodynamically most stable [Figure~\ref{figure2}(g)]. Unlike O and S coverages, nitrogen introduces states at the Fermi level for most of the coverages studied here except of 0.33 and 0.5 ML (Supporting Information). These are found to be semiconducting with 1.40 and 1.66 eV gap, respectively. For 0.5 ML coverage, the VBM and CBM align properly with $\Delta_{\rm VBM}$ = $-$0.07 eV and $\Delta_{\rm CBM}$ = 0.36 eV. In contrast, for 0.33 ML, the band edges align only at pH 9 solution with $\Delta_{\rm VBM}$ = $-$0.07 eV and $\Delta_{\rm CBM}$ = 0.10 eV.

\begin{figure}[t!]
\begin{center}
\includegraphics[angle=90, scale=0.28]{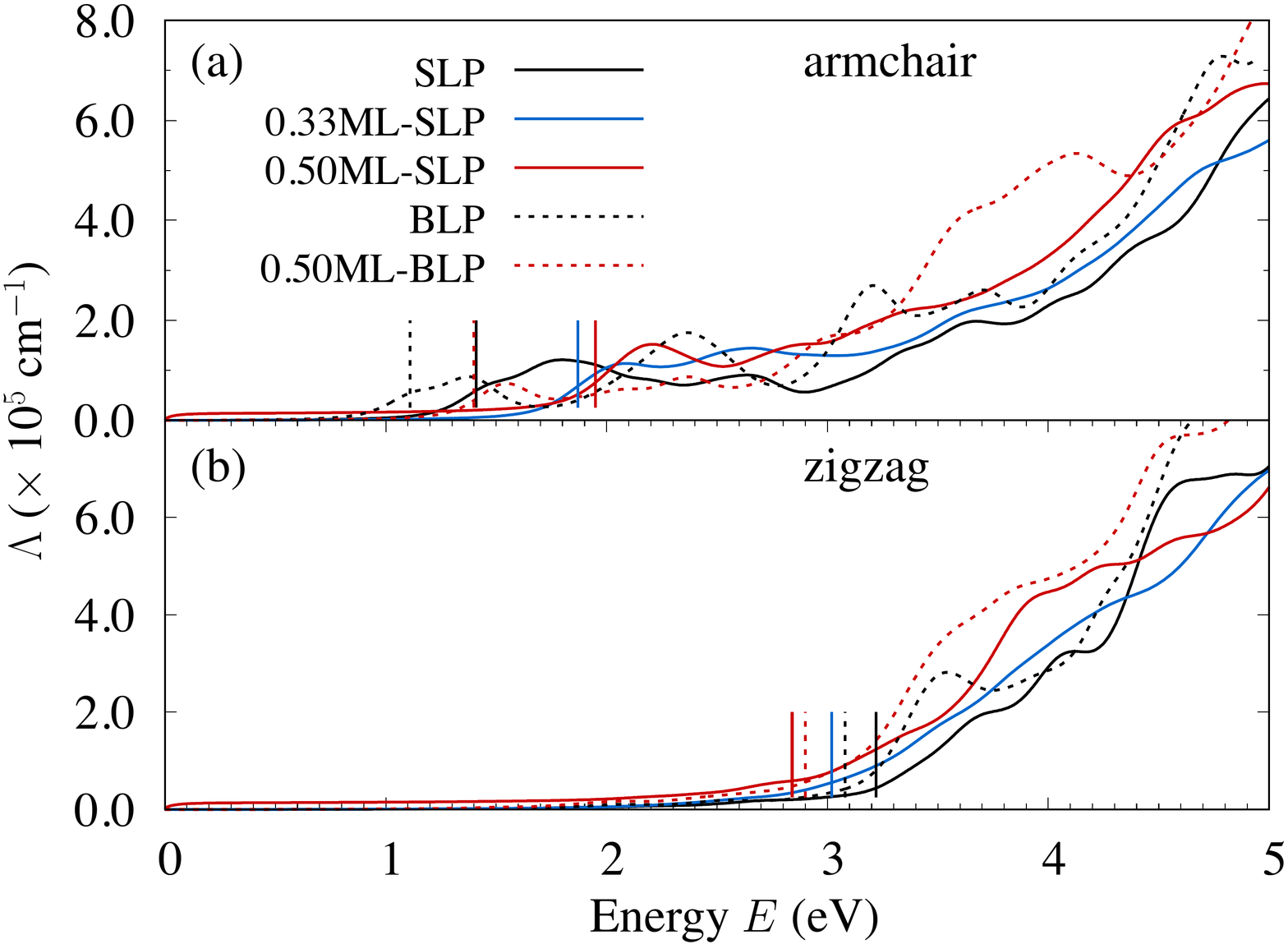}
\caption{Optical absorption coefficients $\Lambda$ for linearly polarized incident light along the (a) armchair, and (b) zigzag direction calculated with HSE06 hybrid exchange-correlation functional. Pristine SLP and BLP are compared with different oxygen coverages. The vertical lines correspond to the absorption edge calculated from the Tauc plots. Anisotropic nature of optical absorption is retained with oxygen coverage along with very high absorption coefficient $\Lambda$ $\sim$ 10$^5$ cm$^{-1}$ in the visible optical range. While the absorption edge is shifted to higher energy for the light polarization along the armchair direction, it is reversed for the incident light polarization along the zigzag directions.  
}
\label{figure_optics}
\end{center}
\end{figure}

\subsection{Optical and excitonic properties}
The frequency dependent complex dielectric tensor, $\varepsilon (\omega) = \varepsilon{'} (\omega) + i \varepsilon{''} (\omega)$, has been used to calculate the optical properties. The imaginary part $\varepsilon{''} (\omega)$ of the linear dielectric tensor is calculated in the long-wavelength $\mathbf{q} \rightarrow \mathbf{0}$ limit within the random phase approximation,~\citep{PhysRevB.73.045112}
\begin{eqnarray}
\varepsilon''_{\alpha \beta}(\omega) & = & \frac{4\pi^2e^2}{\Omega} \lim\limits_{q\rightarrow 0}\frac{1}{q^2}\sum\limits_{c,v,\mathbf{k}}2w_{\mathbf{k}} \delta(\epsilon_{c\mathbf{k}} - \epsilon_{v\mathbf{k}} - \omega) \nonumber \\
&\times& \langle u_{c\mathbf{k}+\mathbf{e}_{\alpha}q} | u_{v\mathbf{k}} \rangle \langle u_{c\mathbf{k}+\mathbf{e}_{\beta}q} | u_{v\mathbf{k}} \rangle^*, \nonumber
\end{eqnarray}
where the factor 2 inside the summation accounts for the spin degeneracy, $\Omega$ is the volume of the primitive cell, and $\omega_{\mathbf k}$ are $k$-point weights. The $\epsilon_{c\mathbf{k}}$ ($\epsilon_{v\mathbf{k}}$) are ${\mathbf k}$-dependent conduction (valence) band energies, $u_{c\mathbf{k}, v\mathbf{k}}$ are cell periodic part of the pseudo-wave-function, and $\mathbf{e}_{\alpha, \beta}$ are unit vectors  along the Cartesian directions. The corresponding real part $\varepsilon{'} (\omega)$ is calculated using Kramers-Kronig transformation.  The absorption co-efficient $\Lambda_{\alpha \alpha} (\omega) = \frac{2\omega}{c}[|\varepsilon_{\alpha \alpha} (\omega)| - \varepsilon_{\alpha \alpha}'' (\omega)]^{\frac{1}{2}}$ is calculated using the hybrid HSE06 exchange-correlation functional.

The electronic structure of phosphorene is anisotropic along different crystallographic directions. Thus, a strong anisotropic optical absorption is expected while the incident light is linearly polarized along the armchair and zigzag directions of phosphorene (Figure~\ref{figure_optics}).  Pristine SLP and BLP show strong dichroism. For the incident light polarized along the armchair direction, the absorption peak is formed at the bandgap (1.41 eV for SLP), which show a sharp layer dependence (1.10 eV for BLP). In contrast, for the light polarized along the zigzag direction, the optical absorption peak is found at $\sim$ 3.22 eV for SLP, which does not show strong layer dependence (3.14 eV for BLP). The overall trend is in good agreement with previous reports.~\citep{10.1038/ncomms5475}  The anisotropic nature of optical absorption remains intact upon oxygen coverage, and both SLP and BLP show similar features upon oxygen coverage. The absorption edge shifts to higher energy with increasing coverage for the light polarization along the armchair direction [Figure~\ref{figure_optics}(a)]. In contrast, the same shifts to lower energy for light polarization along the zigzag direction [Figure~\ref{figure_optics}(b)].  Further, the high absorption coefficient is unaffected by the oxygen coverage reported in Figure~\ref{figure_optics} and remains $\sim$ 10$^{5}$ cm$^{-1}$ for visible light absorption.  

For S coverage, the optical properties show similar qualitative trends that are discussed for oxygen coverage (Figure~\ref{figure_optics}), and shown in the Supporting Information. The absorption remains anisotropic in armchair and zigzag directions. For S coverage with $\Theta \leqslant$ 0.50, the absorption edge is blue shifted for the incident light polarization along the armchair direction, as the band gap increases compared to pristine SLP (1.66 and 1.86 eV for 0.33 and 0.5 ML coverages, respectively). In contrast, the absorption is red shifted for light polarization along the zigzag direction with the absorption edge at 3.05 and 2.90 eV for 0.33 and 0.5 ML coverages, respectively. For both cases that are conducive to water oxidation (Figure~\ref{figure3}),  the absorption coefficient is found to be very high 10$^5$ cm$^{-1}$  within visible light excitation (Supporting Information).

Next, we investigate the excitonic properties in phosphorene derivatives, which are conducive to water splitting. The exciton binding, the bound state energy of an electron-hole pair, is an important metric for photocatalytic efficiency. However, a rigorous treatment within many-body theory with Bethe-Salpeter equation is practically impossible due to exceptional computational cost. In contrast, here we use a simplified hydrogenic model for anisotropic two-dimensional materials, which has been proved to provide results with very reasonable accuracy.\citep{PhysRevB.88.045318, PhysRevB.91.245421, PhysRevLett.116.056401} The parameters in this model, effective hole and electron mass along different crystallographic directions, and dielectric constant at zero frequency are calculated using HSE06 exchange-correlation functional. The obtained results for pristine SLP and BLP are in excellent agreement with previous GW-BSE results~\cite{PhysRevB.89.235319,PhysRevLett.115.066403,PhysRevB.92.165406} and experimental measurement.~\cite{10.1038/nnano.2015.71} The calculated exciton binding $E_b^{\rm ex}$ is 0.76 eV, which confirm strongly bound exciton in SLP. Further, the exciton extension, $\mathcal{R}^{\rm ex}_{\rm arm}$ and $\mathcal{R}^{\rm ex}_{\rm zig}$, is found to be anisotropic with 12.20 and 4.88 \AA, along the armchair and zigzag directions, respectively. Impurity coverage affects both exciton binding and extension. The exciton binding increases in phosphorene with impurity coverage, and further increases with increasing coverage. For example, $E_b^{\rm ex}$ = 0.87 and 0.92 eV for 0.33 and 0.5 ML O-coverage, respectively; whereas the same for S-coverage are found to be 0.90 and 1.13 eV, respectively. In contrast to pristine SLP, the excitons become more localized in these systems,   $\mathcal{R}^{\rm ex}_{\rm arm}$ = 9.73 \AA~ and $\mathcal{R}^{\rm ex}_{\rm zig}$ = 3.99 \AA~ for 0.5 ML O-coverage, while the same for 0.5 ML S-coverage are calculated to be 6.94 and 5.34 \AA, respectively. On the other hand, with increasing layer thickness in BLP,  $E_b^{\rm ex}$ decreases to 0.48 eV, which is 0.75 eV for the 0.5 ML O-BLP system. A large exciton binding may lead to higher recombination and detrimental to photocatalytic application. On the other hand, in these complexes, the very large electron/hole mobility with concurrent effective mass may lead to quick charge separation, and could be consumed in the respective catalytic reaction.  

\section{Conclusions}
In the context of metal-free photocatalyst, a comprehensive investigation of electronic and optical properties of phosphorene derivatives is performed within first-principles calculations. The N, O and S adatoms bind strongly on the two-dimensional phosphorene, and preferentially bind at the dangling site. Further, the binding energy show strong coverage dependence, and 0.25--0.5 ML coverages are found to be most stable. The differential contribution of induced strain due to impurity loading has been discussed. Impurity binding does not introduce midgap impurity states, and the composite system remains semiconducting. At certain coverages (0.33 -- 0.5 ML), which are also thermodynamically stable, the band edges align with the redox potentials for water splitting.  Moreover, the absorption coefficient is found to be very high 10$^5$ cm$^{-1}$ within the visible absorption range. Further, the layer dependence of band alignment and optical properties are investigated in bilayer phosphorene, and the same is found to be capable of oxidation and reduction reactions at 0.5 ML oxygen coverage. Further, we investigate the excitonic properties of the composites conducive to water splitting and compared with the pristine phosphorene.  While two-dimensional phosphorene has been already proposed for potential applications in electronic and optoelectronic applications, the current investigation open up the possibility of using phosphorene derivatives as metal-free photocatalyst.~\citep{ADMA:ADMA201605776} We hope the current study will simulate further theoretical and experimental investigations to take a complete advantage of its outstanding potential.

AS is grateful to UGC, Government of India for financial support in the form of research fellowship. We acknowledge the supercomputing facilities at the Centre for Development of Advanced Computing, Pune; Inter University Accelerator Centre, Delhi; and at the Center for Computational Materials Science, Institute of Materials Research, Tohoku University. M. K. acknowledges the funding from the Department of Science and Technology, Government of India under Ramanujan Fellowship, and Nano Mission project SR/NM/TP-13/2016.


%
\end{document}